\documentclass[prl,twocolumn]{revtex4-1}

\usepackage{amsmath}
\usepackage{mathtools} 
\usepackage{amssymb}
\usepackage{amsthm}
\usepackage{bm,upgreek} 
\usepackage{upgreek} 
\usepackage{xcolor}
\usepackage{graphicx}
\usepackage{epstopdf}
\usepackage[colorlinks,linkcolor=blue,anchorcolor=blue,citecolor=blue,urlcolor=blue]{hyperref}
\usepackage{tikz}
\usepackage[compat=1.1.0]{tikz-feynman}

\newcommand{\bea}{\begin{eqnarray}}
\newcommand{\eea}{\end{eqnarray}}
\newcommand{\ie}{\textit{i.e.{ }}}
\newcommand{\eg}{\textit{e.g.{ }}}

\newcommand{\etal}{\textit{et al.{}}}

\begin{document}


{\bf Reply to “Comment on ‘Anisotropic Scattering Caused by Apical Oxygen Vacancies in Thin Films of Overdoped High Temperature Cuprate Superconductors’ by H. U. Ozdemir \etal”}.  In Ref.~\cite{prl}, we proposed that the apical oxygen vacancies act as anisotropic scattering impurities. Within the Born approximation, this leads to a quasi-particle scattering rate that is maximal (zero) in the antinodal (nodal) direction. This unique angular dependence provides a straightforward mechanism for some puzzling experimental results in overdoped La$_{2-x}$Sr$_x$CuO$_4$ (LSCO) films regarding the superfluid density $\rho_s$ and optical conductivity $\sigma_1$. \cite{Bozovic,Armitage}.


In a recent comment \cite{comment}, the importance of the nature of the impurity scattering is re-emphasized, but some challenges to our picture are raised: that we did not consider the change of the electrostatic potential for in-plane electrons once the apical oxygen is missing (i), the change of Fermi surface topology as the van Hove point is passed (ii), self-energy corrections caused by d-wave pairing (iii), and vertex corrections caused by forward scattering (iv). These concerns are interesting but are either irrelevant or further enhances our conclusions.
We discuss these points one by one below.

Regarding (i), as pointed out in Ref.~\cite{comment}, the change of the electro-static potential is of the order of $50$meV on copper and $3$meV on nearby oxygen. This energy scale is minute in comparison to the local Hubbard interaction on copper (about $10.5$eV) and also to the charge transfer gap (about $4$eV) related to the in-plane oxygen \cite{parameter}.
As a result, this can barely change the effective parameters for the Zhang-Rice singlet \cite{Zhang-Rice}, which serves as the effective degrees of freedom in doped cuprates. In contrast, the hopping integral between the $p_z$-orbital (of the apical oxygen) and the $p_x$/$p_y$ orbitals (of in-plane oxygens) should be of the same order of magnitude as that between the nearest $p_x$ and $p_y$ orbitals (about $0.65$eV \cite{parameter}). As such, the main effect of the apical oxygen vacancy is the depletion of the indirect hopping of the Zhang-Rice singlet via the apical $p_z$-orbital. It is this effect that leads to the anisotropic scattering we discussed in Ref.~\cite{prl}. Nevertheless, we have added an isotropic scattering rate in our study which was shown to be smaller than the anisotropic one by examining the experimental data of superfluid density and optical conductivity.

Regarding (ii), we were aware of the change of the Fermi surface topology with doping, and of the strong correlation nature below optimal doping. This is why we emphasized that we are concerned with the overdoped regime, where the Fermi surface topology no longer changes, and where the normal state could be well described, effectively, by the Landau Fermi liquid picture. In addition, the crucial effect of the anisotropic scattering rate (caused by apical oxygen vacancies) is its vanishment in the nodal direction, which is the key to understand the temperature dependence of the superfluid density, while the Lifshitz transition does not change this essential behavior at all. So this point is irrelevant to our work.

\begin{figure}
\includegraphics[width=0.5\linewidth]{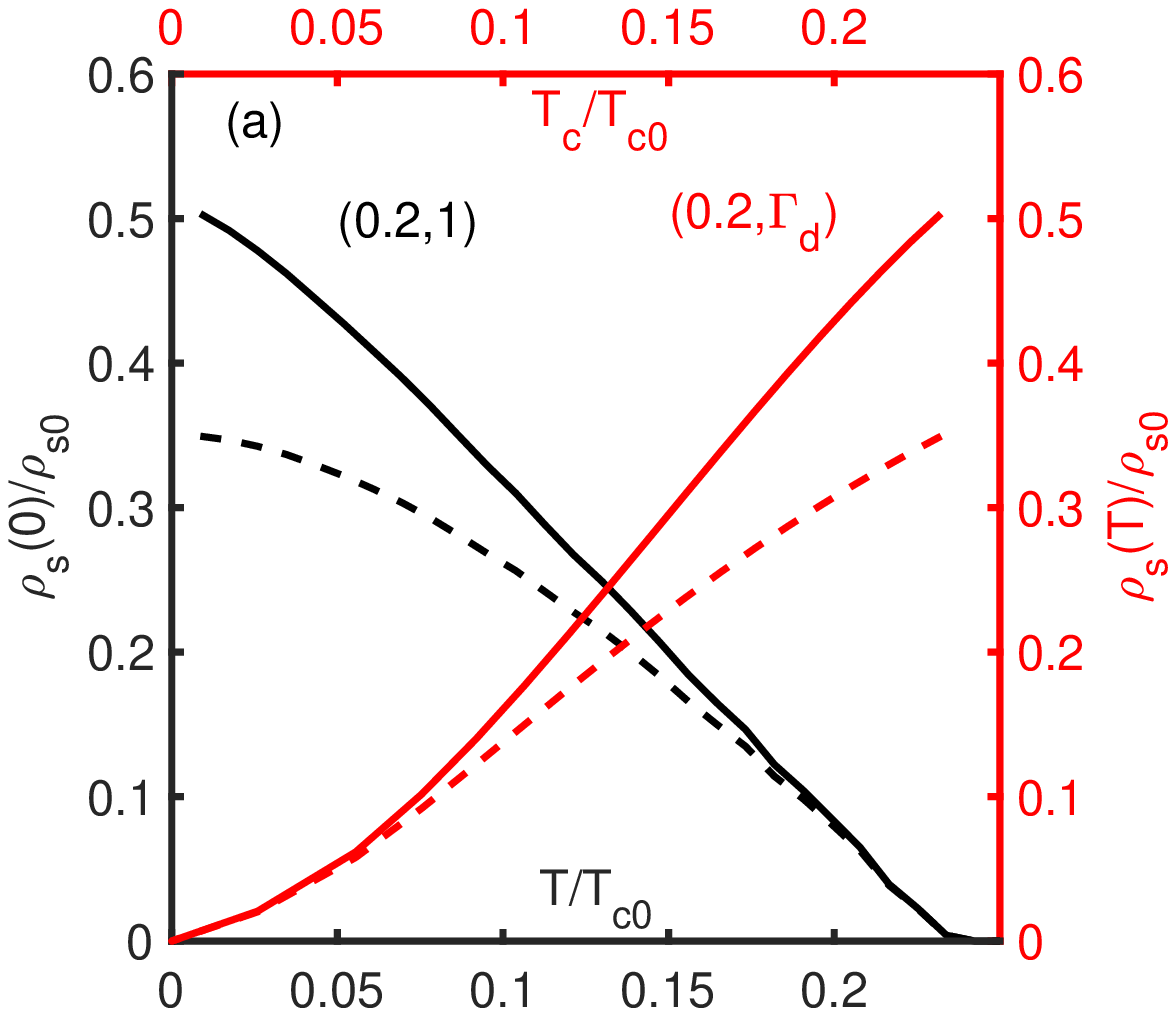}
\quad
\includegraphics[width=0.45\linewidth]{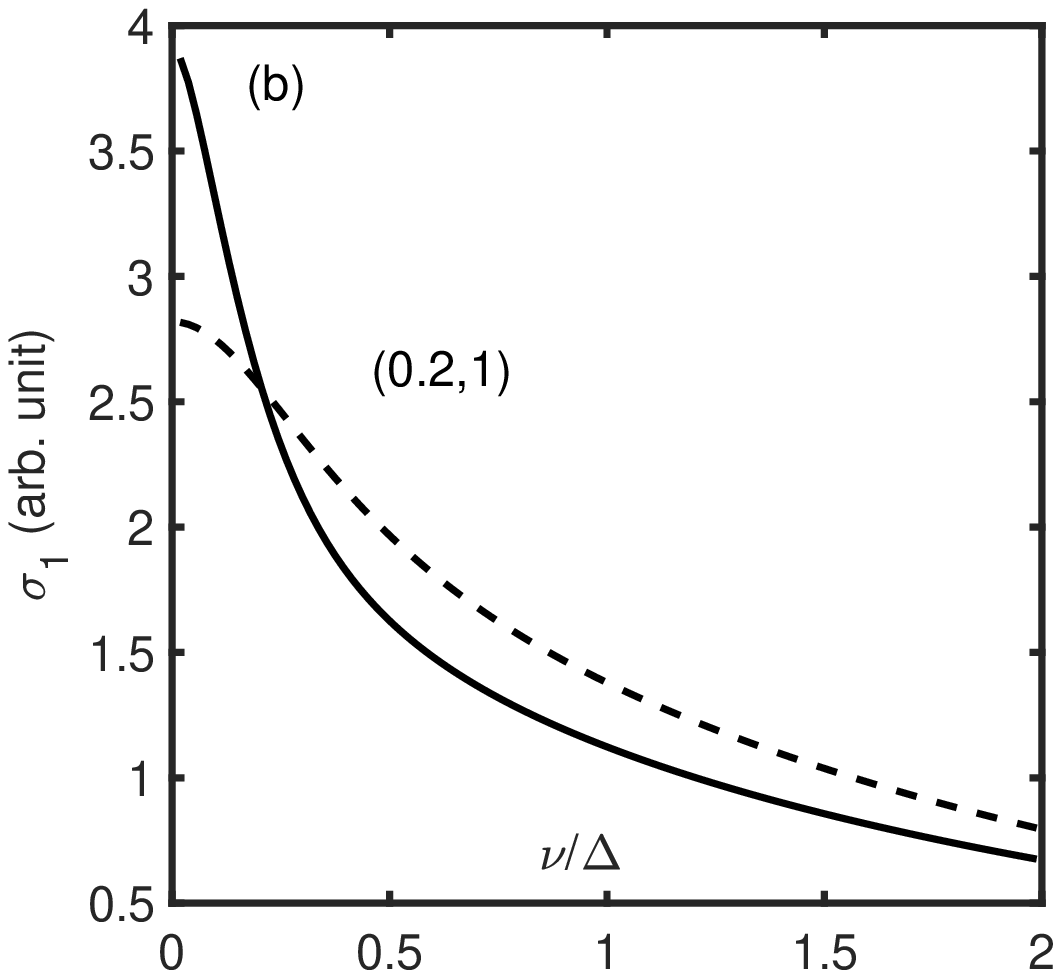}
\caption{\label{fig}
The superfluid density (a) and optical conductivity (b) with (solid lines) or without (dashed lines) self-energy effect. The associated scattering rate is indicated as $(\Gamma_s, \Gamma_d)$ (in unit of $0.87T_{c0}$).}
\end{figure}

Regarding (iii), the self-energy correction altered by d-wave pairing plays important roles in the clean limit ($\Gamma\ll\Delta$), but can be neglected in the dirty limit ($\Gamma\gg\Delta$). In overdoped LSCO films, the scattering rate $\Gamma$ is estimated to be $5\sim10$meV \cite{Armitage}, comparable with or larger than the superconducting gap (estimated by $\Delta\approx2T_c<8$meV), placing the material far away from the clean limit and approaching the dirty limit as $T_c\to0$. We have performed fully self-consistent calculations of the self-energy
\begin{figure}[h]
\begin{align*}
\vcenter{\hbox{
\begin{tikzpicture}
    \begin{feynman}
    \vertex (i) ;
    \vertex[right=0.5cm of i] (a);
    \vertex[right=1cm of a] (b);
    \vertex[right=0.5cm of b] (f);
    \vertex[right=0.5cm of a] (c);
    \node[above=0.5cm of c, crossed dot] (d);
    \diagram*{
        (a) -- [fermion] (b),
        (a) -- [scalar] (d) -- [scalar] (b),
    };
    \end{feynman}
    \end{tikzpicture}}}
~+~
\vcenter{\hbox{
\begin{tikzpicture}
    \begin{feynman}
    \vertex (i) ;
    \vertex[right=0.5cm of i] (a);
    \vertex[right=1cm of a] (b);
    \vertex[right=0.5cm of b] (f);
    \vertex[right=0.5cm of a] (c);
    \diagram*{
        (a) -- [fermion] (b) ,
        (a) -- [photon, half left, looseness=2] (b),
    };
    \end{feynman}
    \end{tikzpicture}}}
\end{align*}
\end{figure}

\noindent under the Born approximation, where the dashed and wavy lines stand for impurity scattering and d-wave BCS pairing interaction, respectively. As $\Delta\to0$, \eg $T\to T_c$ or $T_c\to0$, the self-energy is the same as that of the normal state, leaving only the scattering rate as we studied in Ref.~\cite{prl}. For general cases, we show a group of typical results of the superfluid density and optical conductivity in Fig.~\ref{fig} in the intermediate regime with $\text{max}(\Gamma_s,\Gamma_d)\sim\Delta$. The results are not changed qualitatively, and quantitatively even further enhances our main conclusions: $\rho_s(0)-\rho_s(T)\propto T$ at low temperature, $\rho_s(0)\propto T_c$, and sharpening of $\sigma_1(\nu)$ at low frequencies.

Regarding (iv), we have shown that there is no current vertex correction caused by forward scattering in our model \cite{prl}. In general, the vertex correction caused by forward scattering is captured by the following ladder diagrams,
\begin{figure}[h]
\begin{align*}
\vcenter{\hbox{\begin{tikzpicture}
  \begin{feynman}
    \vertex (i);
    \vertex [below=1.5cm of i] (j);
    \vertex [below=0.75cm of i,crossed dot] (k1);
    \node [right=0.1cm of k1, crossed dot] (k);
    \vertex [right=0.75cm of k1] (v);
    \vertex [right=0.5cm of v] (o);
    \vertex [below=0.25cm of i] (x);
    \vertex [right=0.25cm of x] (i2);
    \vertex [above=0.25cm of j] (y);
    \vertex [right=0.25cm of y] (j2);
    \diagram*{
      (i) --[fermion] (i2) -- [fermion] (v) -- [fermion] (j2) --[fermion] (j),
      (v) --[photon] (o),
      (j2) -- [scalar] (k) -- [scalar] (i2),
    };
  \end{feynman}
\end{tikzpicture}}}
~+~
\vcenter{\hbox{\begin{tikzpicture}
  \begin{feynman}
    \vertex (i);
    \vertex [below=2cm of i] (j);
    \vertex [below=1cm of i,crossed dot] (k1);
    \node [right=0.1cm of k1, crossed dot] (k);
    \node [right=0.45cm of k1, crossed dot] (k2);
    \vertex [right=1cm of k1] (v);
    \vertex [right=0.5cm of v] (o);
    \vertex [below=0.25cm of i] (x);
    \vertex [right=0.25cm of x] (i2);
    \vertex [above=0.25cm of j] (y);
    \vertex [right=0.25cm of y] (j2);
    \vertex [below=0.6cm of i] (x1);
    \vertex [right=0.6cm of x1] (i3);
    \vertex [above=0.6cm of j] (y1);
    \vertex [right=0.6cm of y1] (j3);
    \diagram*{
      (i) --[fermion] (i2) -- [fermion] (i3) -- [fermion] (v) -- [fermion] (j3) -- [fermion] (j2) -- [fermion] (j),
      (v) --[photon] (o),
      (j2) -- [scalar] (k) -- [scalar] (i2),
      (j3) -- [scalar] (k2) -- [scalar] (i3),
    };
  \end{feynman}
\end{tikzpicture}}}
~+~\cdots
\end{align*}
\end{figure}

\noindent which gives the ``transport scattering rate'' $\tau_{\rm tr,k}^{-1}\propto\int_{k'}|V_{kk'}|^2(1-\cos\theta_{k'})$ \cite{AG}.
The impurity scattering matrix $V_{kk'}$ can be decomposed into different symmetry channels, \ie $V_{kk'}=\sum_nV_nf_k^nf_{k'}^n$ with $f_k^n$ the form factor of the $n$-channel. Clearly, if $|V_{kk'}|^2$ does not contain the $p$-wave component (with form factor $\cos\theta_k$), the forward scattering contribution to $\tau_{\rm tr,k}^{-1}$ is zero. In fact, such a conclusion can be obtained more directly and can be generalized to the superconducting state. In the above vertex corrections, if $f_k^n$ are time reversal even for all $n$, as in our case with $f_k^s=1$ and $f_k^d=(\cos k_x-\cos k_y)/2$, the current vertex correction vanishes exactly, as already explained in the supplementary material for Ref.~\cite{prl}.

\vspace{5pt}
D. Wang, J.-Q. Xu, H.-J. Zhang and Q.-H. Wang

National Laboratory of Solid State Microstructures $\&$ School of Physics, Nanjing University, Nanjing 210093, China

Collaborative Innovation Center of Advanced Microstructures, Nanjing 210093, China


\begin{thebibliography}{9}
\bibitem{prl} D. Wang, J.-Q. Xu, H.-J. Zhang and Q.-H. Wang, \href{https://link.aps.org/doi/10.1103/PhysRevLett.128.137001}{Phys. Rev. Lett. {\bf 128}, 137001 (2022)}.
\bibitem{Bozovic} I. Bozovic, X. He, J. Wu and A. T. Bollinger, \href{http://www.nature.com/nature/journal/v536/n7616/full/nature19061.html}{Nature {\bf 536}, 309 (2016)}.
\bibitem{Armitage} F. Mahmood, X. He, I. Bozovic and N. P. Armitage, \href{https://link.aps.org/doi/10.1103/PhysRevLett.122.027003}{Phys. Rev. Lett. {\bf 122}, 027003 (2019)}.
\bibitem{comment} H. U. Ozdemir, V. Mishra, N. R. Lee-Hone, X. Kong, T. Berlijn, D. M. Broun and P. J. Hirschfeld, \href{https://arxiv.org/abs/2206.01301}{arXiv:2206.01301}.
\bibitem{parameter} M. S. Hybertsen, M. Schluter and N. E. Christensen, \href{https://journals.aps.org/prb/abstract/10.1103/PhysRevB.39.9028}{Phys. Rev. B {\bf 39}, 9028 (1989)}.
\bibitem{Zhang-Rice} F. C. Zhang and T. M. Rice, \href{http://link.aps.org/doi/10.1103/PhysRevB.37.3759}{Phys. Rev. B {\bf 37}, 3759 (1988)}.
\bibitem{AG} A. A. Abrikosov and L. P. Gor'kov, Sov. Phys. JETP {\bf 35}, 1090 (1959).
\end{thebibliography}
\end{document}